# Generalized redundancies for time series analysis

Dean Prichard[1]
*Department of Physics,
University of Alaska, Fairbanks, AK 99775*

James Theiler
*Santa Fe Institute, 1660 Old Pecos Trail, Santa Fe, NM 87505;
and Center for Nonlinear Studies and Theoretical Division,
Los Alamos National Laboratory, Los Alamos, NM 87545*

(Draft: May 25, 1994)

## ABSTRACT

Extensions to various information theoretic quantities used for nonlinear time series analysis are discussed, as well as their relationship to the generalized correlation integral. It is shown that calculating redundancies from the correlation integral can be more accurate and more efficient than direct box counting methods. It is also demonstrated that many commonly used nonlinear statistics have information theory based analogues. Furthermore, the relationship between the correlation integral and information theoretic statistics allows us to define "local" versions of many information theory based statistics; including a local version of the Kolmogorov-Sinai entropy, which gives an estimate of the local predictability.

## 1 Introduction

The idea of viewing a chaotic dynamical system as an information source was first suggested by Shaw [1]. Since that time many authors have proposed methods to characterize strange attractors, based on information theoretic quantities. These include information dimension [2,3], various measures of information production rate like the Kolmogorov-Sinai entropy [1,4], as well as its generalizations [5–7] based on the Renyi entropies [8], and information based measures of dependence, such as mutual information [1,9] and redundancies [10,11].

Information theoretic measures are only one of many tools which are available for characterizing nonlinear systems. One class of methods, which includes many of the information based statistics, relies only on the invariant measure of the attractor, this includes various measures of dimension [3,7,12–17] and the statistics [18–23] based on the correlation integral of Grassberger and Procaccia [13]. There are also methods which use dynamical information directly, such as the Lyapunov exponents [24–26], Kolmogorov-Sinai (K-S) entropy, nonlinear prediction error [27–31] and less direct measures of determinism [32–35]. Many of the methods based on the invariant measure can also measure dynamical properties, if a delay

---

[1]Current address: MS-B213, Complex Systems Group, Theoretical Division, Los Alamos National Laboratory, Los Alamos, NM 87545

coordinate embedding is used, however, these methods are not fundamentally dynamical like the Lyapunov exponents, K-S entropy and prediction error are. There is also a growing class of measures which depend on the topological properties of the attractor [36–39].

Many of these methods are related; for example, the Lyapunov exponents are related to the Kolmogorov-Sinai entropy through the Pesin identity [4], and the information dimension and Lyapunov exponents seem to be related through the Kaplan-Yorke conjecture [40]. Further, properties of the unstable periodic orbits, which form the basis of many of the topological methods, can also be used to estimate dimension and entropies [41–45].[2]

The relationship between the Shannon [2] and Renyi entropies [8] and the generalized correlation integral [7] has been pointed out by a number of authors [46–48]. It is demonstrated that by using this relationship, more accurate estimates of many information theoretic statistics can be obtained, as compared to conventional box-counting methods. It is also shown that there are advantages to using statistics based on the generalized entropies of Renyi instead of the Shannon entropy. Furthermore, information theory based analogues to a number of statistics [18,21,22] are proposed. Using the relationship between the Shannon entropy and the generalized correlation integral, "local" versions of many information based statistics are proposed, which include measures of the local coupling between variables; as well as a localized version of the Kolmogorov-Sinai (K-S) entropy, which is related to the local predictability, much like with the local Lyapunov exponents [49–51].

## 2 Entropy, mutual information and redundancies

In this section the definitions of various statistics based on information theory are reviewed. One of the most basic statistics is the Shannon entropy [2], which quantifies the average information gained from a measurement. This is usually estimated from a time series by a box counting approach; that is, a partition size $\delta$ is chosen, and the data $x$ (We use $x$ as a convenient abbreviation for $x(t)$, $t = 1, \ldots, N$) are discretized into integers $y = 1, \ldots, M$ depending on what bin of size $\delta$ they fall into. In this case, the Shannon entropy is given by:

$$H_1(x, \delta) = H_1(y) = - \sum_{y=1}^{M} p(y) \log_2[p(y)]. \tag{1}$$

where $p(y)$ is the probability of being in the $y^{th}$ bin. The actual value of the entropy estimated by this method depends on the partition size $\delta$. For two time series $x_1$ and $x_2$ the joint entropy is given by:

$$H_1(x_1, x_2, \delta) = H_1(y_1, y_2) = - \sum_{y_1} \sum_{y_2} p(y_1, y_2) \log_2[p(y_1, y_2)] \tag{2}$$

and for $m$ variables the entropy (often called block entropy) is:

$$H_1(x_1, \ldots, x_m, \delta) = H_1(y_1, \ldots, y_m) = - \sum_{y_1, \ldots, y_m} p(y_1, \ldots, y_m) \log_2[p(y_1, \ldots, y_m)]. \tag{3}$$

One can also define an entropy for continuous variables:

$$H_1(\vec{x}) = - \int p(\vec{x}) \log_2[p(\vec{x})] d\vec{x} \tag{4}$$

---

[2]It should be pointed out that these methods do not use the linking of the orbits, like the topological methods do.

where $\vec{x} = (x_1, \ldots, x_m)$. However, this definition has some unusual properties, for one it depends on the coordinate system which is used. For example, if $\vec{z} = a\vec{x}$ then $H(\vec{z}) = H(\vec{x}) + \log(a)$, also $\lim_{\delta \to 0} H(\vec{x}, \delta) \neq H(\vec{x})$. The fact that the entropy depends on the partition size allows one to define an information dimension [3], based on the average scaling of the amount of information require to specify a point in the state space within an accuracy of $\delta$. The information dimension is given by:

$$D_1 = \lim_{r \to 0} \frac{-H(\vec{x}, \delta)}{\log \delta} \tag{5}$$

The average amount of information that $x_1$ contains about $x_2$ can be expressed by the mutual information:

$$I_1(x_1; x_2, \delta) = H_1(x_1, \delta) + H_1(x_2, \delta) - H_1(x_1, x_2, \delta). \tag{6}$$

The mutual information is a measure of many bits one can predict about $x_2$ given a measurement of $x_1$ with an accuracy of $\delta$. If $x_1$ and $x_2$ are independent the the mutual information is zero, while if $x_2$ is completely dependent on $x_1$ then $I_1(x_1; x_2) = H_1(x_2)$, also note that $I_1(x_1; x_2) = I_1(x_2; x_1)$ and $I_1(x_1, x_1) = H_1(x_1)$. The mutual information for continuous variables is coordinate independent, unlike the entropy, and assuming that there is a small amount of noise in the data $\lim_{\delta \to 0} I_1(x_1; x_2, \delta) = I_1(x_1; x_2)$. This is because for $\delta$ smaller than the noise scale, both $H_1(x_1, \delta)$ and $H_1(x_2, \delta)$ will scale as $-\log \delta$, while $H(x_1, x_2, \delta)$ will scale as $-2\log \delta$. For a noiseless deterministic system one can use the scaling of the mutual information with $\delta$ to define a "mutual information dimension" [52].

The $m$ dimensional extension of mutual information is called redundancy [10],

$$R_1(x_1; \ldots; x_m, \delta) = \sum_{i=1}^{m} H_1(x_i, \delta) - H_1(x_1, \ldots, x_m, \delta) \tag{7}$$

where $\vec{x}(t) = (x_1(t), x_2(t), \ldots, x_m(t))$ can be either a multivariate signal or a time delay embedding [53] $\vec{x}(t) = (x(t), x(t-\tau), \ldots, x(t-(m-1)\tau))$; for delay coordinates we have:

$$R_1(x_1; \ldots; x_m, \delta) = mH_1(x_1, \delta) - H_1(x_1, \ldots, x_m, \delta). \tag{8}$$

To quantify the amount of information about $x_m$ contained in $x_1, x_2, \ldots, x_{m-1}$ a quantity called marginal redundancy ($R'$) is used [10]:

$$R'_1(x_1, \ldots, x_{m-1}; x_m, \delta) = R_1(x_1; \ldots; x_m, \delta) - R_1(x_1; \ldots; x_{m-1}, \delta). \tag{9}$$

If $x_m$ is independent of $x_1, x_2, \ldots, x_{m-1}$ then the marginal redundancy is zero, while if $x_m$ is completely dependent on $x_1, x_2, \ldots, x_{m-1}$ then $R'_1(x_1, \ldots, x_{m-1}; x_m) = H_1(x_m)$.

The redundancies and marginal redundancies, like the mutual information, are only scale independent for noisy systems. However the quantity

$$\lim_{m \to \infty} [H_1(x_1, \ldots, x_m, \delta) - H_1(x_1, \ldots, x_{m-1}, \delta)] \tag{10}$$

has the opposite behavior, that is, for a deterministic system it is scale independent (both terms scale as $-D_1 \log \delta$, for small $\delta$, so the overall expression does not depend on $\delta$), while for a noisy system it scales as $-\log \delta$. A related quantity is the Kolmogorov-Sinai (K-S) entropy, which is a measure of the mean rate of information creation by the system. Given a time delay embedding $\vec{x}(t) = (x_1(t), x_2(t), \ldots, x_m(t)) = (x(t), x(t-\tau), \ldots, x(t-(m-1)\tau)$, the K-S entropy is:

$$
\begin{aligned}
K_1 &= \lim_{m \to \infty} [H_1(x_m | x_1, \ldots, x_{m-1}, \delta)]/\tau \\
&= \lim_{m \to \infty} [H_1(x_1, \ldots, x_m, \delta) - H_1(x_1, \ldots, x_{m-1}, \delta)]/\tau.
\end{aligned}
\quad (11)
$$

The K-S entropy can also be related to the marginal redundancy [10]; for small delay times $\tau$ we have,

$$
\lim_{m \to \infty} R'_1(\tau) = H_1(x_1) - \tau K_1. \quad (12)
$$

This suggests another way to estimate the entropy is:

$$
K_1 \approx \lim_{m \to \infty} \frac{R'_1(\tau = t_1) - R'_1(\tau = t_2)}{t_2 - t_1}. \quad (13)
$$

## 3  Linear redundancies

Paluš *et al.* [11] define "linear redundancies" which are derived from the continuous case of above formulas for the special case of a multivariate gaussian distribution

$$
p(\vec{x}) = \frac{|\xi_{ij}|^{1/2}}{(2\pi)^{m/2}} e^{-\frac{1}{2} \sum_{i,j}^m \xi_{ij} x_i x_j} \quad (14)
$$

where $|\cdot|$ is the determinant, and $\xi_{ij}$ is an element of the matrix $\boldsymbol{\xi}$, which is the inverse of the covariance matrix $\boldsymbol{\Xi}$, with elements given by: $\Xi_{ij} = \langle (x_i(t) - \langle x_i(t) \rangle)(x_j(t) - \langle x_j(t) \rangle) \rangle$. Combining Eq. (14) and Eq. (4) we find that the "linearized" entropy is given by:

$$
\begin{aligned}
\mathcal{H}_1(\vec{x}) &= \log_2 \left( \frac{(2\pi)^{m/2}}{|\xi_{ij}|^{1/2}} \right) + \frac{1}{2 \log_e(2)} \int \left( \sum_{i,j}^m \xi_{ij} x_i x_j \right) p(\vec{x}) d\vec{x} \\
&= \frac{m}{2} \log_2(2\pi e) + \frac{1}{2} \log_2 |\Xi_{ij}|
\end{aligned}
\quad (15)
$$

and the linearized redundancy[3] is:

$$
\mathcal{R}(\vec{x}) = \frac{1}{2} \sum_i^m \log_2(\Xi_{ii}) - \frac{1}{2} \log_2 |\Xi_{ij}|. \quad (16)
$$

This is equivalent to the form given in Paluš *et al.* [11] since for a symmetric matrix $|\boldsymbol{\Xi}| = \prod_j^m \lambda_j$ where $\lambda_j$ are the eigenvalues of $\boldsymbol{\Xi}$. Paluš *et al.* also define what they call a marginal linear redundancy by:

$$
\mathcal{R}'(x_1, x_2, \ldots, x_{m-1}; x_m) = \mathcal{R}(x_1; x_2; \ldots; x_m) - \mathcal{R}(x_1; x_2; \ldots; x_{m-1}). \quad (17)
$$

Computing the linear redundancies provides a way of assessing the role of linear correlations in the estimate of the actual information-theoretic quantity. If the redundancy and linear redundancy are substantially different, then there is substantial nonlinearity in the time series. Thus, one has a qualitative test for nonlinearity. In some cases it can be advantageous to transform the original data to have a gaussian distribution, so a nongaussian distribution is not mistaken for nonlinearity. Paluš [54] has also proposed combining this test with the method of surrogate data [55]. One computes the redundancies and their linearized versions for the original data set, as well as for and ensemble of surrogate data sets which are generated to match the linear properties (the power spectrum) of the original data set. If the redundancies for the original data are significantly different from the values for the surrogates, then one can formally reject the null hypothesis that the data arise from a linear process. It is also possible to obtain a quantiative statement about the confidence level of the evidence for nonlinearity. Thus, the comparisons with linear surrogate data and the comparisons with a linear statistic provide complementary information about the possible nonlinearity in the time series. Further, as pointed out by Paluš, comparing the linear redundancies calculated from the original and surrogate data sets gives a good way to check for that the surrogate data sets really are reproducing the linear properties of the original data.

---

[3]One can also define nonlinear statistics based on "local" versions of the "linearized" statistics.

## 4 Relation to $C_1$

The most straightforward way to estimate the quantities defined above is to use a box counting approach: the $m$ dimensional space is divided into a number of boxes of size $\delta$. By counting the number of points $n_i$ in the $i^{\text{th}}$ box, the probability can be estimated as $p_i \approx n_i/N$ where $N$ is the total number of points.[4] A number of authors have used refinements to this procedure, by adapting the size of the boxes depending on the local density [9–11].

It has been shown by Liebert and Schuster [46] that $H_1(\vec{x}, r)$ can be related to $C_1(\vec{x}, r)$, the generalized correlation integral of order 1. Instead of estimating probabilities $p(\vec{x}, \delta)$ within boxes of size $\delta$, one can calculate probabilities $P(\vec{x}, r)$ in regions of radius $r$ about each point. The two are related by:

$$\sum_{\substack{i \\ \text{(bins)}}} p_i(\vec{x}, \delta) \log_2[p_i(\vec{x}, \delta)] = \frac{1}{N} \sum_{\substack{t \\ \text{(datapoints)}}} \log_2[p_{i(t)}(\vec{x}, \delta)] \approx \frac{1}{N} \sum_t \log_2[P_t(\vec{x}, r)] = \log_2 C_1(\vec{x}, r) \tag{18}$$

where $p_i(\vec{x}, \delta)$ is the probability of being in the the $i^{\text{th}}$ bin (a box of diameter $\delta$), $i(t)$ is the box that the $t^{\text{th}}$ data point is in, and $P_t(\vec{x}, r)$ is the probability of being in the box of radius $r$ (or diameter $\delta = 2r$) centered at the point $\vec{x}(t)$. Since the two boxes are the same size and very close to each other (they overlap), we can heuristically justify the relation $p_i(\delta) \approx P_t(r)$ for $\delta = 2r$ (using maximum norm).[5]

A natural estimate[6] of $P_t(r)$ is given by $B(\vec{x}(t), r)$, which is the fraction of data points

---

[4] For small $n_i$ Grassberger [56] has derived a correction to the formula, which for the Shannon entropy is $p_i \log_2[p_i] \approx \frac{n_i}{N} \left( \log_2(N) - \Psi(n_i) - \frac{(-1)^{n_i}}{n_i+1} \right)$ where $\Psi(x)$ is the digamma function (see also Wolpert and Wolf [57]).

[5] For maximum norm, a radius $r$ corresponds to a box size (diameter) $\delta = 2r$. For the euclidean norm, in an $m$ dimensional space, a sphere of radius $r$ has the same volume as a cube of diameter $\delta = c_m^{-1/m} r$, where $c_m$ is the the volume of a m-dimensional unit sphere.

[6] Grassberger [56] has also derived a small $n_t$ correction to this formula, $\log_2 P_i \approx$

(excluding $\vec{x}(t)$ itself) within $r$ of the $\vec{x}(t)$. That is,

$$P_t(\vec{x}, r) \approx B(\vec{x}(t), r) = \frac{n_t}{N} = \frac{1}{N} \sum_{\substack{s \\ s \neq t}} \Theta(r - \|\vec{x}(t) - \vec{x}(s)\|) \qquad (19)$$

where $\Theta$ is the Heaviside function, $n_t$ is the number of points within a radius $r$ of $\vec{x}(t)$, and $\|\cdot\|$ is some measure of distance (we use maximum norm). The generalized correlation integral of order 1 is a (geometric) average of the $B(\vec{x}(t), r)$ probabilities, and is given by

$$\log C(\vec{x}, r) = \frac{1}{N} \sum_t \log B(\vec{x}(t), r). \qquad (20)$$

Note that Eq. (19) is a simple form of kernel density estimation [58], often this would be written in the form:

$$P_t(\vec{x}, r) \approx \frac{1}{N} \sum_{\substack{s \\ s \neq t}} K\left(\frac{\|\vec{x}(t) - \vec{x}(s)\|}{r}\right) \qquad (21)$$

where $K(z) = 1$ if $z < 1$ and otherwise is zero. This particular kernel is far from optimal (in fact, Silverman [58] calls it the "naive estimator"). However, even this crude form of kernel density estimation is generally considered superior to using a multidimensional histogram (binning). Better results can often be obtained, if one uses a kernel $K(z)$ which decreases with increasing $z$ or even a kernel whose width depends on the local density (see Ref. [58] for details). However, in this paper Eq. (19) will be used, so we can estimate the Shannon entropy from the generalized correlation integral of order one [7, 16, 46]

$$H_1(x_1; x_2; \ldots; x_m, \delta) \approx -\log C_1(x_1, x_2, \ldots, x_m, r). \qquad (22)$$

We can now express the redundancies (Eqs 7 and 9) and K-S entropy (Eqs. 11 and 13) in terms of $C_1$, and since kernel density estimation is being used instead of binning, we can expect more accurate results with limited data sets.

## 5 Generalized entropies and redundancies

Instead of using the Shannon entropy to calculate redundancies, we can generalize these statistics by using the Renyi entropies [8]:

$$H_q(\vec{x}, \delta) = \frac{1}{1-q} \log_2 \sum_i [p_i(\vec{x}, \delta)]^q. \qquad (23)$$

It is easy to show that the limit as $q \to 1$ leads to the Shannon entropy. Again we can relate probabilities $p(\vec{x}, \delta)$ within boxes of size $\delta$, to probabilities $P(\vec{x}, r)$ in regions of radius $r = \delta/2$ (for maximun norm) about each point:

$$\sum_i [p_i(\vec{x}, \delta)]^q \approx \frac{1}{N} \sum_t [P_t(\vec{x}, r)]^{q-1} \qquad (24)$$

---

$\left(\Psi(n_t + 1) - \log_2(N) - \frac{(-1)^{n_t}}{n_t + 1}\right)$, which is used in the calculations below.

so we have[7],

$$H_q(\vec{x}, \delta) \approx \frac{1}{1-q} \log_2 \left( \frac{1}{N} \sum_t \left( \frac{1}{N} \sum_{\substack{s \\ s \neq t}} \Theta(r - \|\vec{x}(t) - \vec{x}(s)\|) \right)^{q-1} \right) = -\log_2[C_q(\vec{x}, r)] \quad (25)$$

where $C_q(\vec{x}, r)$ is the generalized correlation integral [7, 16]. The idea of relating the Renyi entropies to the generalized correlation integral is by no means new. It was mentioned in the review by Grassberger *et al.* [47], the $q = 1$ case was used by Liebert and Schuster [46] and Pompe [48] used the correlation integral to calculate $H_2$. However, by using Eq. (25) we can define generalized redundancies $R_q(\vec{x}, r)$ and $R'_q(\vec{x}, r)$ in terms of $C_q(\vec{x}, r)$. Pompe [48] has also proposed what he calls a generalized mutual information which he expresses in terms of the second order ($q = 2$) correlation integral. Pompe's generalized mutual information is given by:

$$Q_2(x_1, \ldots, x_m) = H_2(x_m) - H_2(x_1, \ldots, x_m) + H_2(x_1, \ldots, x_{m-1}) \quad (26)$$

which is the same as the second order ($q = 2$) generalized marginal redundancy.

While $q = 1$ leads to the natural definition of the entropy[8] there are reasons to prefer different values of $q$. For example, for all $q$ except $q = 2$ there are corrections to Eq. (24) for small $r$. Grassberger has derived the asymptotic form of these corrections [56], but for finite length sets, the best statistics, at small $r$, are obtained by using $q = 2$ (In contrast, for box counting methods there are small $n_i$ corrections for all $q$ [56]). Another reason for using $q = 2$ is that it is the fastest to compute of all the generalized correlation integrals. Further, the correlation integral has a dynamic range of $O(N^2)$ as opposed to $O(N)$ for box counting methods, this permits the use of smaller values of $r$ [17]. Finally, the speed of box counting methods [59] is not an issue, as there are numerous fast correlation integral algorithms available [60–62] so one can compute the generalized redundancies for small $r$ in $O(N \log N)$ time or faster.

### 5.1 Generalized linear redundancies

We can also define linear versions of the generalized entropies,

$$\mathcal{H}_q(\vec{x}) = \frac{1}{(1-q)} \log_2 \left( \frac{|\xi_{ij}|^{q/2}}{(2\pi)^{qm/2}} \int e^{-\frac{q}{2} \sum_{i,j}^m \xi_{ij} x_i x_j} d\vec{x} \right) \quad (27)$$

using $|q\xi_{ij}| = q^m |\xi_{ij}|$ and the normalization condition of the gaussian we find:

$$\begin{aligned}
\mathcal{H}_q(\vec{x}) &= \frac{1}{(1-q)} \log_2 \left[ \frac{|\xi_{ij}|^{(q-1)/2}}{q^{m/2}(2\pi)^{(q-1)m/2}} \right] \\
&= \frac{m}{2} \log_2(2\pi) + \frac{1}{2} \log_2 |\Xi_{ij}| + \frac{m \log_2(q)}{2(q-1)},
\end{aligned} \quad (28)$$

---

[7]One can also express the probability in $m$ dimensions as $P = k/c_m r_m(k)^m$ where $r_m(k)$ is the distance to the $k^{\text{th}}$ nearest neighbor, and $c_m$ is the volume of a "sphere" of radius $r$; it depends both on the embedding dimension and the distance norm),so it should be possible to make "fixed mass" versions of the generalized redundancies as well.

[8]That is, $q = 1$ is the only one of the generalized entropies which is an additive quantity.

and for the generalized linear redundancies:

$$\mathcal{R}(\vec{x}) = \frac{1}{2} \sum_i^m \log_2(\Xi_{ii}) - \frac{1}{2} \log_2 |\Xi_{ij}| \tag{29}$$

which is the same as Eq. (16); that is, the linear redundancies do not depend on $q$, but the "linear entropies" (Eq. (28)) do depend on $q$, as does the linearized version of the generalized correlation integral

$$\mathcal{C}_q(\vec{x}) = - \left[ \frac{|\xi_{ij}|^{(q-1)/2}}{q^{m/2}(2\pi)^{(q-1)m/2}} \right]^{1/(1-q)}. \tag{30}$$

Therefore, the idea of comparing a nonlinear statistic to its "linearized" version can be extended to all the statistics based on the correlation integral (*e.g.* Refs. [5, 18, 19, 21, 22]).

## 6 Applications

### 6.1 Clean computer generated data

As the first test, the correlation integral based redundancy analysis is applied to 8192 points from the chaotic Rössler equations [63] (with parameters $a = 0.15$, $b = 0.2$, $c = 10$ and a sampling time of $\delta t = 0.314$, these are the same parameters used by Paluš [64]). In Fig. 1 we show the linear, $C_2$, and $C_1$ based redundancies and marginal redundancies as a function of the time delay $\tau$, with $r = 0.1\sigma$, where $\sigma$ is the standard deviation of the data set. The lines are for increasing embedding dimensions ($m = 2, \ldots, 8$) starting at the bottom of the graph. Eq. (12) suggests that as $m$ is increased the marginal redundancy curves as a function of $\tau$ should accumulate to a line which has a slope equal to $-1$ times the $K_q$ entropy. From the slope of the $C_2$ based marginal redundancies (Fig. 1b) it can be seen that the $K_2$ entropy is roughly 0.03 bits/timestep (a similar value is found using the estimate of Eq. (11)). However, a reliable estimate of the $K_1$ entropy can not be obtained with this number of points (see Fig. 1d) using either the method of Eq. (11) or Eq. (12).[9] That is, as suggested above, by using the $C_2$ based redundancies we can either use smaller $r$ or fewer data points. In Fig. 1(e-f) the linear redundancies and marginal redundancies are shown for comparison. Notice the difference between the $\tau$ dependence of redundancies and their linearized versions.

### 6.2 Real data: SFI-A

This method is also applied to 8192 points from a chaotic laser experiment, which exhibits Lorenz-type chaos with a correlation dimension of roughly 2.05 and a positive $K_2$ entropy [65]. This data set (A.cont) was part of the time series competition sponsored by the Santa Fe Institute [66]. In Fig. 2 we show the results of the analysis for the linear, $C_1$, and $C_2$ based redundancies as a function of the time delay $\tau$ and for embedding dimensions $m = 2, \ldots, 8$. For the $C_2$ based redundancies we use $r = 0.1\sigma$, however, when using this value of $r$ for the $C_1$ based redundancies there are positive slopes at large $m$ and $\tau$ for the marginal redundancy, as in figure Fig. 1d. The best results are obtained for $r = 0.25\sigma$, which is what is shown in Fig. 2(c-d). The linear redundancies and marginal redundancies are shown in Fig. 2(e-f) for comparison. The difference between the shapes of the redundancies and their linearized version clearly shows that there is nonlinearity in this data set. From the slopes of the $C_2$ based marginal redundancies it is seen that the $K_2$ entropy is roughly 0.025 bits/timestep (we find $K_2 \approx 0.03$ bits/timestep from Eq. (11)). However, we are unable to get an estimate of the $K_1$ entropy with this number of data points. Paluš [64] also examines this data set and gets very similar results to Fig. 2(c-d) using an adaptive box counting method.

---

[9]The results are slightly better at $r = 0.25\sigma$.

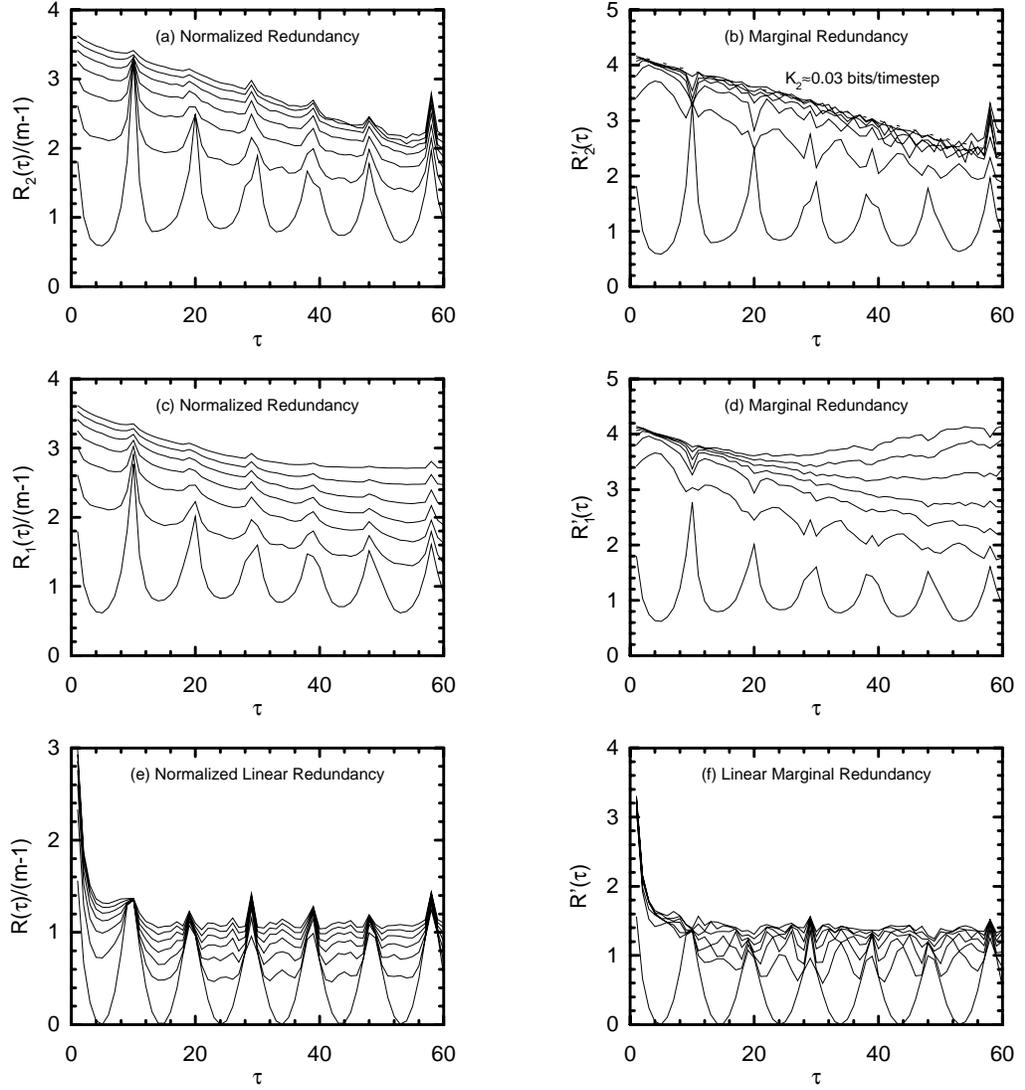

**Fig. 1.** Redundancy analysis for Rossler data set, for embedding dimensions $m = 2-8$ and time delays $\tau = 1-60$ sample times. All curves are for $r = 0.1\sigma$ where $\sigma$ is the standard deviation of the time series. (a) $C_2$ based normalized redundancies ($R_2(\tau)/(m-1)$). (b) $C_2$ based marginal redundancies ($R'_2(\tau)$). (c) $C_1$ based normalized redundancies ($R_1(\tau)/(m-1)$). (d) $C_1$ based marginal redundancies ($R'_1(\tau)$). (e) Normalized linear redundancies ($\mathcal{R}(\tau)/(m-1)$). (f) Linear marginal redundancies ($\mathcal{R}'(\tau)$).

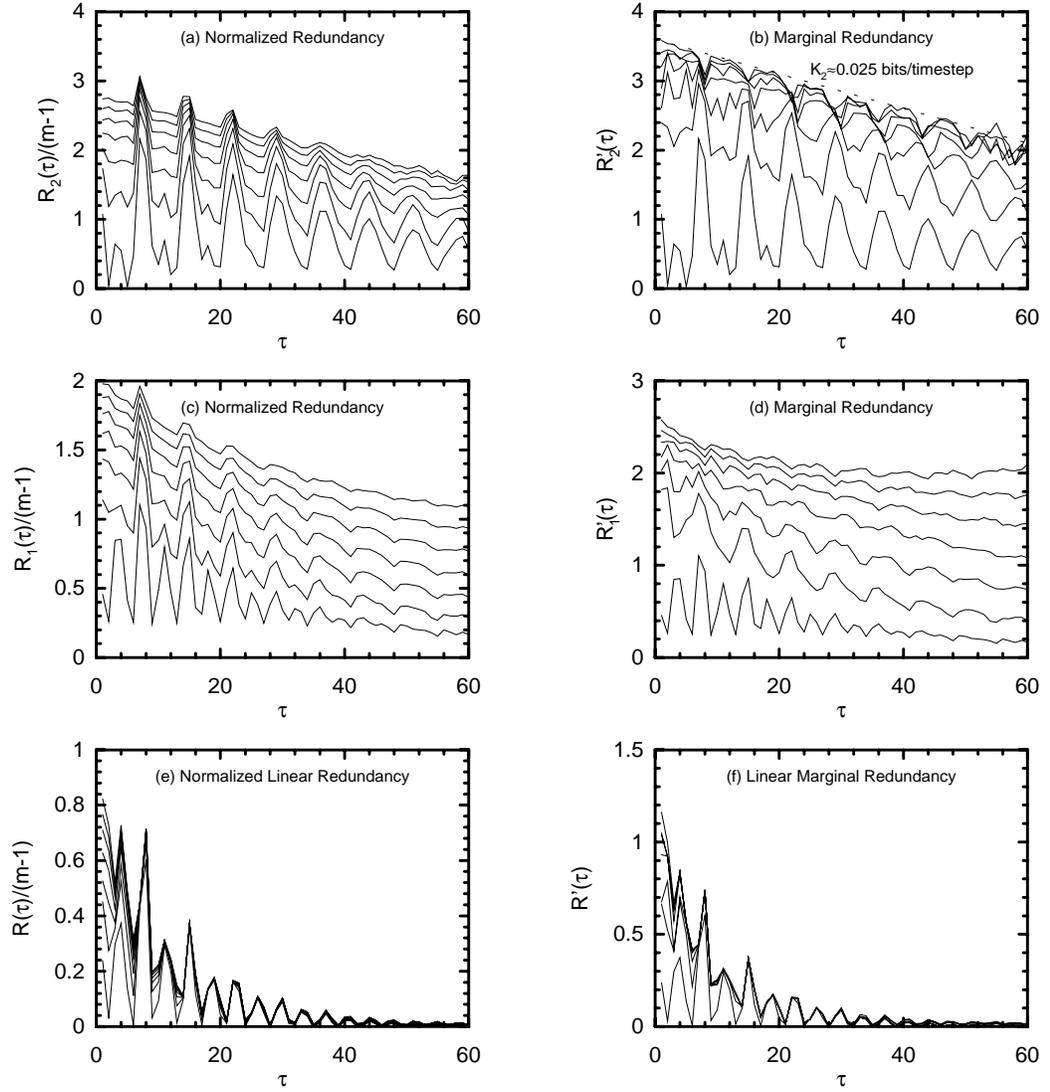

**Fig. 2.** Redundancy analysis for SFI-A data set, for embedding dimensions $m = 2 - 8$ and time delays $\tau = 1 - 60$ sample times. Curves (a) and (b) are for $r = 0.1\sigma$, curves (c) and (d) are for $r = 0.25\sigma$, where $\sigma$ is the standard deviation of the time series. (a) $C_2$ based normalized redundancies ($R_2(\tau)/(m-1)$). (b) $C_2$ based marginal redundancies ($R'_2(\tau)$). (c) $C_1$ based normalized redundancies ($R_1(\tau)/(m-1)$). (d) $C_1$ based marginal redundancies ($R'_1(\tau)$). (e) Normalized linear redundancies ($\mathcal{R}(\tau)/(m-1)$). (f) Linear marginal redundancies ($\mathcal{R}'(\tau)$).

## 7 Relation to other statistics

The connection between the entropy and the correlation integral allows us to relate other correlation integral based statistics to their analogues from information theory. For example, if the sequence is IID then $R_q$ will be zero; this is the idea behind the BDS test [18–20]. Putting $R_q$ in terms of the correlation integral (for delay coordinates) we find:

$$R_q(\vec{x}, r) = \log_2(C_q(\vec{x}, r)) - \log_2([C_q(x_1, r)]^m) \tag{31}$$

which is very similar to the BDS statistic

$$\mathrm{BDS}_q(\vec{x}, r) = C_q(\vec{x}, r) - [C_q(x_1, r)]^m. \tag{32}$$

Green and Savit [22] have also proposed a statistic, based on the correlation integral, to quantify the amount of additional information in $x_m$ about $x_1$ which is not a result the dependence of $x_1$ on $x_2, \ldots x_{m-1}$. One measure of this is given by the conditional redundancy[10]:

$$R_q(x_1; x_m | x_2, \ldots, x_{m-1}, r) = R'_q(x_1, \ldots, x_{m-1}; x_m, r) - R'_q(x_2, \ldots, x_{m-1}; x_m, r) \tag{33}$$
$$= -\log_2\left(\frac{C_q(x_1, \ldots, x_{m-1}, r) C_q(x_2, \ldots, x_m, r)}{C_q(x_1, \ldots, x_m, r) C_q(x_2, \ldots, x_{m-1}, r)}\right).$$

For a time delay embedding this reduces to:

$$R_q(x_1; x_m | x_2, \ldots, x_{m-1}, r) = -\log_2\left(\frac{C_q(x_1, \ldots, x_{m-1}, r)^2}{C_q(x_1, \ldots, x_m, r) C_q(x_1, \ldots, x_{m-2}, r)}\right). \tag{34}$$

The statistic proposed by Green and Savit is:

$$\zeta_q(x_1; x_m, r) = 1 - \left(\frac{C_q(x_1, \ldots, x_{m-1}, r) C_q(x_2, \ldots, x_m, r)}{C_q(x_1, \ldots, x_m, r) C_q(x_2, \ldots, x_{m-1}, r)}\right), \tag{35}$$

which for time delays reduces to the statistic of Savit and Green [21]

$$\delta_q(x_1; x_m, r) = 1 - \left(\frac{C_q(x_1, \ldots, x_{m-1}, r)^2}{C_q(x_1, \ldots, x_m, r) C_q(x_1, \ldots, x_{m-2}, r)}\right). \tag{36}$$

We are not advocating the use of these information theory based statistics in place of the BDS or Green and Savit statistics, but rather, pointing out that there are information theory based analogues to these statistics. An important distinction between these statistics, and statistics like the correlation dimension and the K-S entropy, is that they are evaluated at a fixed $r$ as opposed to taking the limit as $r \to 0$. Another measure of this type is the 'ApEn' statistic advocated by Pincus [23], which is just the $K_2$ entropy [5, 6] evaluated at a fixed $m$ and $r$.

---

[10] We are grateful to Milan Paluš for pointing out that this difference of marginal redundancies can be written as a conditional redundancy.

## 7.1 Cross-redundancies

Estimating the relations between multiple time series is an important problem. Recently statistics have been proposed to estimate nonlinear correlations between variables [22, 67]. One simple measure, which we call the cross-redundancy, is given by:

$$I_q(x_1; x_2, l, r) = H_q(x_1(t), r) + H_q(x_2(t + l), r) - H_q(x_1(t), x_2(t + l), r) \tag{37}$$

where $l$ is a lag time, as in a cross-correlation (A similar statistic has been used by Vastano and Swinney [68] for measuing infomation transport in spatiotemporal systems). That is, the cross-redundancy is just the mutual information between $x_1$ and a lagged value of $x_2$. The cross-redundancy can also be expressed in terms of the correlation integral:

$$I_q(x_1; x_2, l, r) = -\log_2 \left( \frac{C_q(x_1(t), r) C_q(x_2(t + l), r)}{C_q(x_1(t), x_2(t + l), r)} \right). \tag{38}$$

It can be seen that this is very similar to the quantity defined in Eq. (33) (the analog of the statistic of Green and Savit), the important difference being the use of the lag time. As with the other statistics we can define a linearized version of the cross-redundancy:

$$\mathcal{I}_q(x_1; x_2, l) = -\frac{1}{2} \log_2(1 - (\Xi_{x_1 x_2}(l))^2) \tag{39}$$

where it is assumed that both of the series $x_1$ and $x_2$ have zero mean and unit variance, and $\Xi_{x_1 x_2}(l) = \langle x_1(t) x_2(t+l) \rangle^{1/2}$ is the cross-correlation function between $x_1$ and $x_2$ as a function of the lag time $l$.

As an example, the second order ($q = 2$) cross-redundancy and its linearized version are computed for the $x$ and $y$ components of the Lorenz equations [69] (with parameters $\sigma = 10$, $\beta = 8/3$, and $r = 28$, and a sampling time of $\delta t = 0.04$). In Fig. 3 we show the cross-redundancy (solid lines) for $r$ values $0.01\sigma, 0.02\sigma, \ldots, 0.1\sigma$ as a function of lag time (the top curve is $r = 0.01\sigma$), and the linear cross-redundancy (dashed curve). Notice both the cross-redundancy and its linearized version show a peak at a lag of roughly $-2$ time steps, but that the cross-redundancy also shows (nonlinear) correlations at longer lags, which are not detected by the linear cross-redundancy. That is, linearly $y$ looks just like a lagged version of $x$, but by using the cross-redundancy it is seen that there is nonlinear coupling between $x$ and $y$.

## 7.2 Local information based measures

The relationship between the correlation integral and the entropy makes it easy to define "local" information theoretic measures (local in state space), for example, the local version of the Shannon entropy is:

$$h_1(\vec{x}(t), r) = -\log_2 \left( \frac{1}{N} \sum_{\substack{s \\ s \neq t}} \Theta(r - \|\vec{x}(t) - \vec{x}(s)\|) \right) = -\log_2(B(\vec{x}(t), r)). \tag{40}$$

We can also define local redundancies, based on the inner sum of the correlation integral ($B(\vec{x}(t), r)$), as well as a local version of the K-S entropy:

$$\begin{aligned} k_1(\vec{x}(t)) &= \lim_{m \to \infty} [h_1(x_1(t), \ldots, x_m(t), r) - h_1(x_1(t), \ldots, x_{m-1}(t), r)]/\tau \\ &= \lim_{m \to \infty} \frac{1}{\tau} \left[ \log_2 \left( \frac{B(x_1(t), \ldots, x_{m-1}(t), r)}{B(x_1(t), \ldots, x_m(t), r)} \right) \right] \end{aligned} \tag{41}$$

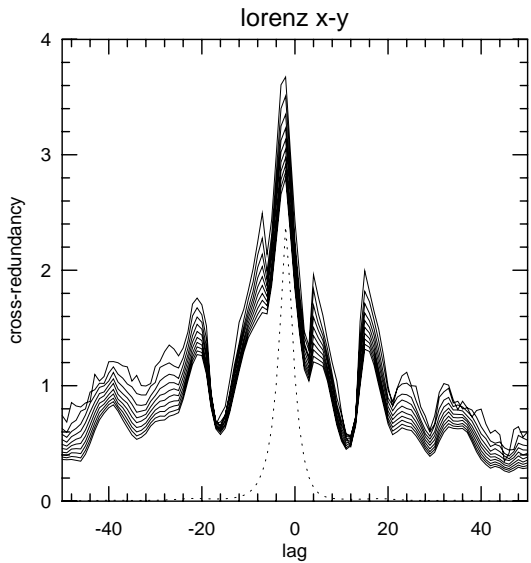

**Fig. 3.** Cross-redundancy between the $x$ and $y$ components of the Lorenz equations as a function of time lag (solid curves). Top curve is for $r = 0.01\sigma$, next is for $r = 0.02\sigma$ and so on to $r = 0.1\sigma$. Dashed line is for the linear cross-redundancy.

where $(x_1(t), x_2(t), \ldots, x_m(t))$ is a time delay embedding: $(x(t), x(t - \tau), \ldots, x(t - (m - 1)\tau)$. The local K-S entropy gives us a measure of the local predictability, without actually doing nonlinear prediction, or calculating local Lyapunov exponents [49–51][11]. If one is just interested in determining how the degree of predictability changes across the state space, we suggest that calculating the local entropy may be a numerically easier way to get this information that using nonlinear prediction or local Lyapunov exponents.

As an example of the local K-S entropy we generate $N = 65536$ points of the $x$ and $z$ variables of the Rössler equations with a time step $\delta t = 0.314$. The local entropy is then estimated in embedding dimensions $m = 3, \ldots, 8$ using the $x$ component, a time delay of $\tau = 5$ (the first minimum of the mutual information) and $r = \sigma/4$ for the first 500 points along the trajectory. (Since we are only interested in finding how the predictability changes in different regions of state space, and not in the exact value of the local entropy, we do not take the limits as $r \to 0$ and $m \to \infty$, but instead evaluate the approximate local entropy at finite $r$ and $m$). By examining the Rössler attractor it is clear that most of the stretching and folding occurs when $z$ is large, therefore, we expect the local entropy to be large only when $z$ is large. In Fig. 4 the approximate local entropy for embedding dimensions $m = 3, \ldots, 8$ is shown, as well as the $z$ component of the Rössler equations. Notice that the local entropy has "spikes" when $z$ is large, as was expected. In the figure the curves are shifted to the right by $\tau = 5$ for each increasing embedding dimension, because of the time delay embedding.

Minimizing mutual information is one criteria that might be used to get a good embedding, in fact, this was suggested by Shaw and explored by Fraser [9]. Casdagli *et al.* [70]

---

[11] The local K-S entropy is related to the local Lyapunov exponents of Eckhardt [51]. The "local" exponents of Abarbanel *et al.* [49,50] are in fact finite time Lyapunov exponents, that is they are more like finite time averages of Eq. (41).

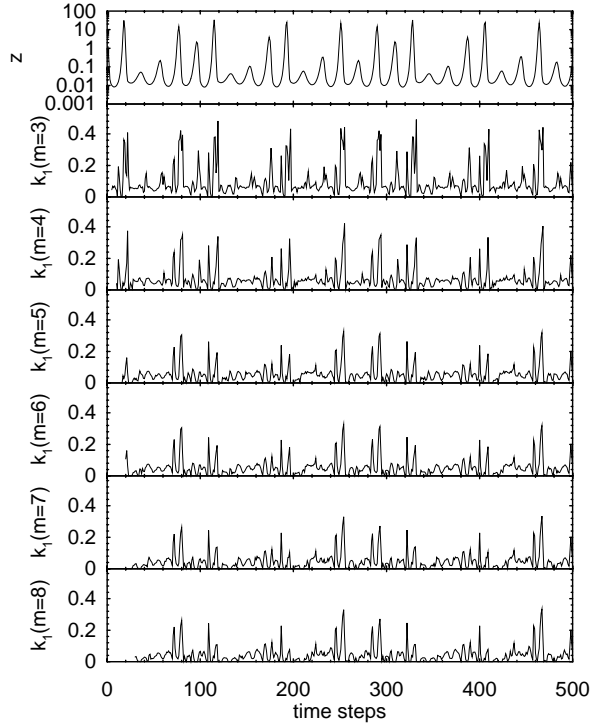

**Fig. 4.** Top panel: $z$ component of the Rössler equations. Next 6 panels are the (approximate) local entropy calculated from the $x$ component for embedding dimensions $m = 3, \ldots, 8$.

suggest minimizing another quantity which they call the "conditional variance"

$$\mathrm{Var}(z|\vec{y}) = \int z^2 p(z|\vec{y})dz - \left(\int zp(z|\vec{y})dz\right)^2. \qquad (42)$$

For delay coordinates, $z = x(t+\tau)$ and $\vec{y} = (x(t), \ldots, x(t-(m-1)\tau))$. Casdagli *et al.* [70] argue that conditional variance is a more appropriate criteria than mutual information. Another benefit of the conditional variance is that is a local measure. Čenys *et al.* [67, 71] have also proposed statistics based on the conditional variance.

Casdagli *et al.* also point out that the quality of the embedding depends on the coupling between the variables, they illustrate this idea using the Lorenz equations:

$$\begin{aligned} \dot{x} &= \sigma(y-x) \\ \dot{y} &= rx - xz - y \\ \dot{z} &= xy - \beta z \end{aligned} \qquad (43)$$

they point out that when $x$ is small the coupling between $y$ and $z$ is weak, so in the presence of any noise one expects a poor reconstruction when $x \approx 0$. This suggests that we might want to look at a quantity like $i_1(z(t); y(t)|x(t))$ along a trajectory in the state space to determine how coupling between $y$ and $z$ depends on the position in state space. We can express $i_1(z(t); y(t)|x(t))$ in terms of the inner sum of the correlation integral $(B(\vec{x}(t), r))$:

$$i_1(z(t); y(t)|x(t), r) = h_1(y(t), z(t), r) + h_1(x(t), y(t), r) - h_1(x(t), y(t), z(t), r) - h_1(x(t), r)$$

$$= -\log_2 \left( \frac{B(y(t), z(t), r) B(x(t), y(t), r)}{B(x(t), y(t), z(t), r) B(x(t), r)} \right). \tag{44}$$

As an example, we generate $N = 65536$ points of $x$, $y$, and $z$ from the Lorenz equations with a time step of $\delta t = 0.01$. Using $r = 3.25$ (roughly 1/4 of the standard deviation) $i_1(z(t); y(t)|x(t))$ is computed for the first 500 points along the trajectory. In Fig. 5 we show $i_1(z(t); y(t)|x(t))$ versus $x$. Notice that $i_1(z(t); y(t)|x(t))$ is small near $x = 0$, since when $x$ is small the coupling between $y$ and $z$ is weak.

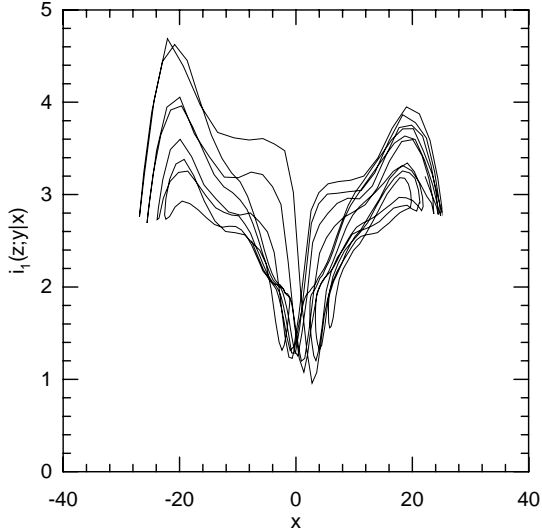

**Fig. 5.** The local mutual information between $z$ and $y$ given $x$, ($i_1(z; y|x)$) plotted against $x$ for a short trajectory of the Lorenz equations. When $x$ is small the coupling between $y$ and $z$ is weak, therefore, $i_1(z; y|x)$ is small.

## 8 Conclusions

We have discussed the relationship between various information theory based quantities and generalized correlation integral of order 1, and extensions of these quantities based on the generalized correlation integral. It has been shown that the correlation integral approach has several advantages over box counting methods (especially for $q = 2$), and that the idea of comparing the $\tau$ dependence of a statistic to its "linearized" version can be extended to all the statistics based on the correlation integral. Finally, we have introduced new information theoretic statistics, including "local" versions of several statistics based on the inner sum of the correlation integral.

## 9 Acknowledgments

We thank Milan Paluš, Thomas Schreiber, Channon Price and John Williams for their comments and suggestions. This work was partially supported by NIMH grant 1-R01-MH47184 and the US Department of Energy.